\documentclass[aps,twocolumn,showpacs]{revtex4}

\usepackage[xdvi]{graphicx}
\usepackage{latexsym}
\usepackage{amsbsy}
\usepackage{amsmath}
\usepackage{amssymb}

\begin{document}

\date{\today}

\title{Current and field driven domain wall motion under influence of the
  Dzyaloshinsky-Moriya interaction}  

\author{R.\ Wieser}
\affiliation{International Center of Quantum Materials,
  Peking University, No. 209 Chengfu Road, Haidian District Beijing,
  100871, China}

\begin{abstract}
A complete analytical description of the dynamics of current and field driven
transverse domain walls under the influence of the
Dzyaloshinsky-Moriya interaction using the $\phi-q$ model will be
given. Five different scenarios will be observed where the
Dzyaloshinsky-Moriya vector is either parallel or perpendicular to the easy
axis anisotropy of the system and a direct reversal respectively
precessional motion will be assumed.
\end{abstract}

\maketitle

\section{Introduction} \label{s:intro}
Due to the usability for data storage
\cite{parkinSCIENCE08,hayashiSCIENCE08}, information 
transport \cite{menzelPRL12} and as logic elements
\cite{allwoodSCIENCE05,khajetooriansSCIENCE11} magnetic wires and spin
chains became more and more important during the last years.  
An important role in all these applications play the motion of domain
walls which underlies some criteria: it shall be fast, controllable and stable. 

While the domain wall motion of a normal transverse domain wall 
is well understood the influence of the Dzyaloshinsky-Moriya
interaction is mostly tuned out. However, the
Dzyaloshinsky-Moriya interaction is induced by the spin-orbit
interaction and appears if the symmetry of the system is broken. This
is the case at surfaces. Now, nearly all magnetic nanowires and spin
chains are deposit on a surface. Therefore, the influence of a
Dzyaloshinsky-Moriya interaction on the domain wall motion should not
be ignored.         

A first overview of the influence of the Dzyaloshinsky-Moriya
interaction on the motion a transverse domain wall has been given in the
pioneering works of Tretiakov and Abanov \cite{tretiakovPRL10}, and
Thiaville et al. \cite{thiavilleEPL12}. However, these 
publications don't cover all possible scenarios. While
\cite{tretiakovPRL10} describes the influence of a Dzyaloshinsky-Moria
vector parallel to the easy axis directions in the case of a current
driven domain wall, \cite{thiavilleEPL12} describes the influence of a
Dzyaloshinsky-Moria vector perpendicular to the easy axis driven by an
external field. Furthermore, in \cite{thiavilleEPL12} the description
is restricted to the direct reversal scenario during the domain wall
motion. The possibility of a precession during the domain wall motion
has been ignored. Therefore, we can say so far there is no publication
which describes the field and current driven domain wall motion under
the influence of the Dzyaloshinsky-Moriya interaction, which covers
all possible scenarios. This leak of information shall be 
closed with this publication. Furthermore, the description given by
Tretiakov and Abanov is quite complex. This publication presents a
simple way which is easy to understand and which offers a deeper
understanding.  
 
The next section introduces the $q-\phi$ model which has been used to
investigate the domain wall motion of a transverse domain wall under
the influence of the Dzyaloshinsky-Moriya interaction.  

\section{$q-\phi$ model} \label{s:model}
A starting point for the analytical description of any kind of domain wall
structures and their dynamics is the Micromagnetic continuum
approximation of the classical Heisenberg model. The magnetic
properties of a 1D or quasi-1D domain wall with influence coming from
the Dzyaloshinsky-Moriya interaction are well described by the
following Heisenberg Hamiltonian: 
\begin{eqnarray} \label{Heisenberg}
{\cal H} &=& -J\sum\limits_{\langle i j \rangle} \vec{S}_i\cdot
\vec{S}_j - \vec{D}_{\mathrm{DM}}\cdot \sum\limits_{\langle i j
  \rangle} \vec{S}_i\times \vec{S}_j - \mu_SB_z 
\sum\limits_{i} {S}_i^z \nonumber \\
&+& D_h \sum\limits_{i} \left(S_i^y\right)^2 - D_e
\sum\limits_{i} \left(S_i^z\right)^2 \;, 
\end{eqnarray} 
where $\vec{S}_i = \vec{\mu}/\mu_S$ are 3D magnetic moments of unit
length. 

The first term of this Hamiltonian describes the ferromagnetic
exchange coupling between nearest neighbors with the coupling constant
$J > 0$. The second term describes the asymmetric exchange (Dzyaloshinsky-Moriya
interaction) between nearest neighbors coming from the spin-orbit
coupling. Depending on the symmetry of the system 
the Dzyaloshinsky-Moriya vector $\vec{D}_{\mathrm{DM}}$ is oriented
perpendicular or parallel to the bonding axis of the neighboring spins
$\vec{S}_i$ and $\vec{S}_j$ \cite{vedmedenkoPRB07}. The third sum describes
the influence of an external magnetic field $B_z$ in $z$ direction and
the last two terms represent uniaxial anisotropies with a hard axis oriented in
$ \pm y$ direction and an easy axis oriented in $\pm
x$ direction. Within this manuscript we will assume 
orientations of the Dzyaloshinsky-Moriya vector
$\vec{D}_{\mathrm{DM}}$ parallel respectively 
perpendicular to the easy axis anisotropy. The precise orientation
will be announced at the beginning of every subsection.

The Hamiltonian Eq.~(\ref{Heisenberg}) can be used to calculate the
dynamics of domain wall numerical
\cite{wieserPRB04,schiebackEPJB07,wieserPRB10ii}. However, a discrete 
structure is not convenient for an analytical description. In these
cases it is comfortable to use the micromagnetic  
continuum description. The corresponding energy: 
\begin{equation}
E = \int\limits_{-\infty}^{+\infty}\!{\cal E}(\theta,\phi)\mathrm{d}z
\end{equation} 
can be got from $\cal H$ by performing a Taylor expansion up to the
first order \cite{brownBOOK63}. Furthermore, it makes sense to use
spherical instead of the cartesian coordinates: $S_x =
\sin\theta\cos\phi$, $S_y = 
\sin\theta\sin\phi$, and $S_z = \cos\theta$. The detailed energy 
densities $\cal E$ corresponding to the different scenarios will be
given in Sec.~\ref{s:Para} and 
\ref{s:Perp}.   

\begin{figure}[h]
\includegraphics*[bb = 40 340 590 785,width=8.cm]{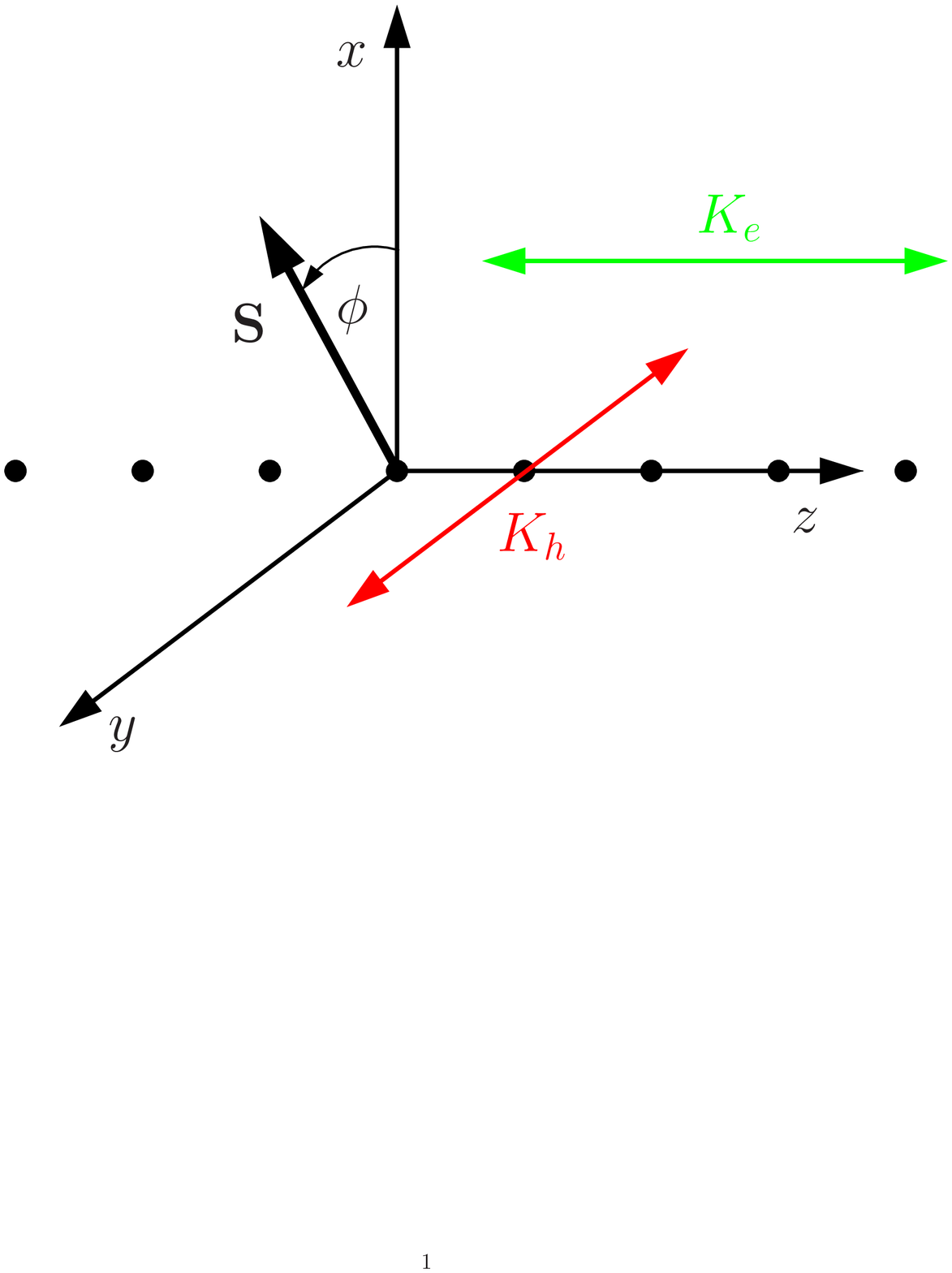}  
  \caption {Underlying coordinate system: $K_e$ and $K_h$ are
  the uniaxial easy respectively hard axis anisotropy of the system. $\phi$ is
  the angle between the magnetic moment $\mathbf{S}$ in the center of
  the domain wall and the $x$ axis; $\theta$ with respect to
  the $z$ axis (not show). The dots symbolizes the next lattice sites
  of the 1D or quasi-1D spin system.}        
  \label{f:pic0}
\end{figure}

Our goal is to give a complete description of the dynamics of a
$180^\circ$ transverse domain wall in a 1D or quasi-1D system with
long axis in $z$ direction. Due to this fact we assume for all
scenarios that the easy axis anisotropy is oriented in $\pm
z$ direction to take into account the contributions from 
crystalline anisotropy as well as dipole-dipole interaction \cite{wieserPRB04}.
Furthermore, if $D_h \neq 0$ (additional hard axis anisotropy) the hard
axis is oriented in $\pm y$ direction (see
Fig.~\ref{f:pic0}). Under these assumptions in all
cases the magnetic moments in the domains are oriented in $\pm
z$ direction and the orientation of the magnetic moment in the center
of the domain wall depends on $D_h$: if $D_h > 0$ the magnetic moment
will be oriented in one of the $x$ directions ($\pm x$, $\phi = 0$ or
$\phi = \pi$), if $D_h = 0$ the symmetry in the $xy$-plane is not
broken and the magnetic  
moment can be oriented in any of the possible directions in the
$xy$-plane ($0 \leq \phi \leq 2\pi$). So far we haven't take into
account the Dzyaloshinsky-Moriya interaction. In the following we
assume that the Dzyaloshinsky-Moriya interaction is small with
respect to the other energy contributions (exchange and anisotropy):
$|\vec{D}_{\mathrm{DM}}| = D_{\mathrm{DM}} \ll J$, $D_{\mathrm{DM}}
\ll D_e$, and $D_{\mathrm{DM}} \ll D_h$. The assumption of a small
Dzyaloshinsky-Moriya interaction is necessary to 
find domain wall solutions instead of spin spirals.

Within the micromagnetic continuum description it is easy to verify
that with the assumptions made before the static solution $(B_z 
= 0)$ of a $180^\circ$ transverse domain wall is given by
\cite{brownBOOK63,malozemoffBOOK79,hubertBOOK98}: 
\begin{eqnarray} \label{thetaMouginDMmotion}
\theta(z,t) = 2 \mathrm{arctan}\left(e^{-\frac{z-q(t)}{\Delta}}\right)\;,
\end{eqnarray}
and
\begin{eqnarray} \label{profile}
S_z = \cos \theta = \pm \tanh\left(\frac{z-q(t)}{\Delta}\right)\;. 
\end{eqnarray}
This profile is the general solution of a 1D or quasi-1D $180^\circ$
domain wall and independent from the explicit form of the Hamiltonian
given by Eq.~(\ref{Heisenberg}), as long as the Dzyaloshinsky-Moriya
interaction is small: $\theta$ is the angle with respect to
the $z$ axis, $\Delta$ the domain wall width, which depends on $\cal
H$, and $q(t)$ is the center 
of the domain wall. The center of the domain wall itself is characterized
by $S_z = \cos\theta = 0$ which means $\theta = \pi/2$. Additional, we
find for the variation $\delta{\cal E}/\delta \theta$ of the
  microscopic energy density ${\cal E}$ under the condition to be in
  the center of the domain wall ($\theta = \pi / 2$):
\begin{equation} \label{BedMougin0DMmotion}
\left.\frac{\delta{\cal E}}{\delta \theta}\right|_{\theta = \pi/2} = M_SB_z\;.
\end{equation}
A detailed description is not necessary at this point and will be
given later in Sec.~\ref{s:Para} and \ref{s:Perp}. 
As before in Eq.~(\ref{Heisenberg}) $B_z$ is the external
field. $M_S = \mu_S / a^3$ is the saturation magnetization, and $a$ is
the lattice constant. Additional we can deduce the following
conditions from the domain wall profile: 
\begin{equation} \label{BedMougin1DMmotion}
\frac{\mathrm{d} \theta}{\mathrm{d} z} = -\frac{\sin \theta}{\Delta} 
\end{equation}
and
\begin{equation}  \label{BedMougin3DMmotion}
v = \frac{\mathrm{d}q}{\mathrm{d} t} = \frac{\Delta}{\sin
  \theta}\frac{\mathrm{d}\theta}{\mathrm{d} t}\;, 
\end{equation}
where $v$ is the velocity of the domain wall.   

The underlying equation of motion is the Gilbert equation with
additional spin torque terms (adiabatic and non-adiabatic) describing 
the influence of an electric current $u_z \propto j$, with $j$ the
current density \cite{schiebackEPJB07,wieserPRB10ii}: 
\begin{eqnarray} \label{GEqS} 
\frac{\mathrm{d}\vec{S}}{\mathrm{d} t} &=& \gamma \vec{S}
\times \mathbf{H}_{\mathrm{eff}} - \alpha
\vec{S} \times \frac{\mathrm{d}\vec{S}}{\mathrm{d} t} \nonumber \\ 
&-& u_z \frac{\mathrm{d}\vec{S}}{\mathrm{d} z} + \beta u_z
\vec{S} \times \frac{\mathrm{d}\vec{S}}{\mathrm{d} z} 
\;.
\end{eqnarray} 
The first term of this equation describe the precessional motion of
the magnetic moment $\vec{S}$ in
the effective field $\mathbf{H}_{\mathrm{eff}} = - (1/\mu_S)(\mathrm{d}\cal
H/\mathrm{d}\vec{S})$. $\gamma$ is the gyromagnetic ratio. The 
second term is the so called Gilbert damping with damping constant
$\alpha$ \cite{gilbertIEEE04}. The third and fourth terms are the adiabatic and
non-adiabatic spin torque terms describing the influence of an
polarized electric current. The adiabatic spin torque term appears due
to the momentum conservation during the interaction between the localized
magnetic moments $\vec{S}$ and the spins of the electrons. The
non-adiabatic spin torque term takes into account the scattering processes of
the electrons. $\beta$ is the so called non-adiabaticity constant. For
a detailed description see e.g. \cite{slonczewskiJMMM96,zhangPRL04}.   
The corresponding micromagnetic equations in spherical coordinates are:
\begin{eqnarray} 
\frac{\mathrm{d}\theta}{\mathrm{d} t} &=&
\frac{\gamma}{M_S\sin\theta}\frac{\delta {\cal E}}{\delta  
\phi}  - \alpha\sin\theta\frac{\mathrm{d}\phi}{\mathrm{d} t} - u_z
\frac{\mathrm{d} \theta}{\mathrm{d} z} -  
\beta u_z \sin\theta  \frac{\mathrm{d} \phi}{\mathrm{d} z} \;, \nonumber
\\ \label{G1DWDM} 
\end{eqnarray}
and
\begin{eqnarray} 
\frac{\mathrm{d}\phi}{\mathrm{d} t} &=&  - \frac{\gamma}{M_S
  \sin\theta}\frac{\delta {\cal E}}{\delta  
\theta} + \frac{\alpha}{\sin\theta}\frac{\mathrm{d}\theta}{\mathrm{d}
  t} - u_z \frac{\mathrm{d} \phi}{ 
\mathrm{d} z} + \frac{\beta u_z}{\sin\theta}\frac{\mathrm{d}
  \theta}{\mathrm{d} z}\;. 
\nonumber \\ \label{G2DWDM}
\end{eqnarray}
$\delta {\cal E}/\delta \theta$ respectively $\delta {\cal E}/\delta
\phi$ are the variations of the energy density with respect to $\theta$
respectively $\phi$.

Inserting the conditions
Eq.~(\ref{BedMougin0DMmotion})-(\ref{BedMougin3DMmotion}), together
with the assumption to be in the center of the domain wall: $\theta =
\pi/2$, lead to the following equations which are the starting point of
our considerations:
\begin{eqnarray} 
\frac{v}{\Delta} &=&
\frac{\gamma}{M_S}\frac{\delta {\cal E}}{\delta  
\phi}  - \alpha\frac{\mathrm{d}\phi}{\mathrm{d} t} + 
\frac{u_z}{\Delta} -  
\beta u_z  \frac{\mathrm{d} \phi}{\mathrm{d} z} \;, \nonumber
\\ \label{Goofy1} 
\end{eqnarray}
and
\begin{eqnarray} 
\frac{\mathrm{d}\phi}{\mathrm{d} t} &=&  - \gamma B_z+
\frac{\alpha v}{\Delta} - u_z \frac{\mathrm{d} \phi}{ 
\mathrm{d} z} - \frac{\beta u_z}{\Delta}\;. 
\nonumber \\ \label{Goofy2}
\end{eqnarray}
The first equation (\ref{Goofy1}) describes the velocity $v$ and the
second equation (\ref{Goofy2}) the precession of 
the domain wall.

These equations are simple to solve and depend only on $q$
(or $v =  \mathrm{d}q/\mathrm{d}t$) and $\phi$. In the next sections
the explicit 
form of $\phi(z,t)$ under the assumptions of a Dzyaloshinsky-Moriya
interaction with a vector $\vec{D}_{\mathrm{DM}}$ parallel and
perpendicular to the easy axis 
anisotropy shall be characterized and the corresponding equations
(\ref{Goofy1}) and (\ref{Goofy2}) solved. 

A detailed outline of the following sections can be found in following
table: the first and second column give the section and sketch
(Fig.) of one scenario and the last four column the
corresponding informations about the 
used assumptions (orientation of $\vec{D}_{\mathrm{DM}}$, $D_e$ and
$D_h$ as well as the reversal 
mechanism where direct means $\mathrm{d}\phi/\mathrm{d} t = 0$ and
precession $\mathrm{d}\phi/\mathrm{d} t \neq 0$). 

\begin{table}[h]
\caption{}
\label{tab1}
\begin{tabular}{|c|c|c|c|c|c|}
\hline
Sec. & Fig. & $D_e$ & $D_h$ & $\vec{D}_{\mathrm{DM}}$ & reversal \\
\hline
\ref{S1D} & \ref{f:pic1} & $\pm z$ & $\pm y$ & $+z$ & direct  \\
\hline
\ref{S1P} & \ref{f:pic2} & $\pm z$ & no anisotropy & $+z$ & precession \\
\hline
\ref{S2D} & \ref{f:pic3} & $\pm z$ & $\pm y$ & $+y$ & direct  \\
\hline
\ref{S3D} & \ref{f:pic4} & $\pm z$ & $\pm y$ & $+x$ & direct \\
\hline
\ref{S2P} & \ref{f:pic5} & $\pm z$ & no anisotropy & $+y$ & precession \\
\hline
\end{tabular}
\end{table}

This table can be seen as a road map of this publication. The calculations
themselves (dealing with main equations Eq.~(\ref{Goofy1}) and
(\ref{Goofy2})) are simple and follow always the same procedure: After writing 
down the energy density $\cal E$, corresponding to ${\cal H}$
[Eq.~(\ref{Heisenberg})] and the made assumptions (see table \ref{tab1}), the
variation $\delta {\cal E}/\delta \phi|_{\theta = \pi/2} = 0$ has to
be calculated. Inserting the result in the main equations
Eq.~(\ref{Goofy1}) and (\ref{Goofy2}) together with an assumption
about the angle $\phi$ lead to the final equations describing the
domain wall motion. Independent of the chosen assumptions $\delta
{\cal E}/\delta \theta = 0$ lead to the domain wall profile 
Eq.~(\ref{profile}) together with the domain wall width $\Delta$.

\section{DM vector parallel to the easy axis anisotropy
  directions} \label{s:Para}  
In the following we assume a spin chain with a transverse domain
wall with an easy axis anisotropy in 
$\pm z$ direction. We will stay with this assumption till the end of
this publication. Furthermore, in this section we assume that the
Dzyaloshinsky-Moriya 
vector is parallel to the easy axis direction: $\vec{D}_{\mathrm{DM}}
= D_{\mathrm{DM}} \vec{e}_z$. Then in the following two subsections the two
possible scenarios $D_h \neq 0$ and $D_h = 0$ will be discussed. In
the first scenario the hard axis anisotropy $D_h$ leads to a symmetry
break and a direct 
reversal of the magnetic moments during the motion. In the
second scenario (no hard axis anisotropy, $D_h = 0$) the system shows
a rotational symmetry around the long axis ($z$ axis).

In both cases the micromagnetic energy density
$\cal E$ is given by: 
\begin{eqnarray} \label{EohneDM}
{\cal E} &=& A \left[\left(\frac{\mathrm{d} \theta}{\mathrm{d} z}  
\right)^2 + \sin^2\theta \left(\frac{\mathrm{d} \phi}{\mathrm{d}
  z}\right)^2\right]  + {\cal D}_{\mathrm{DM}} \sin^2 \theta
\frac{\mathrm{d} \phi}{\mathrm{d} z}  \nonumber \\
&-& M_S B_z\cos \theta  + K_h \sin^2\theta \sin^2\phi
- K_e \cos^2\theta   \;,
\end{eqnarray}
where $A = J/2a$ is the exchange stiffness, $K_\eta = D_\eta a^3$
($\eta \in \{e,h\}$) the uniaxial anisotropies, $B_z$ the external
field, ${\cal D}_{\mathrm{DM}} = D_{\mathrm{DM}}/a^2$ the
micromagnetic Dzyaloshinsky-Moriya interaction, $M_S = \mu_S/a^3$ the
magnetization, and $a$ the 
lattice constant. The difference between both scenarios is the appearance
of the hard axis anisotropy: first scenario $K_h \neq 0$ (direct
reversal), second scenario $K_h = 0$ (precessional motion). 

In the following we will continue with the micromagnetic description
instead of the discrete description used in
Eq.~(\ref{Heisenberg}).
\\    

\subsection{Direct Reversal} \label{S1D} In this subsection we assume
a hard axis anisotropy $K_h \neq 0$ in $\pm y$ direction which is
dominating: $K_h \gg {\cal D}_{\mathrm{DM}}$. This immediately leads
to the fact that we can expect: $\mathrm{d}\phi/\mathrm{d}z = 0$. In
principle due to $K_h \gg {\cal D}_{\mathrm{DM}}$ the
Dzyaloshinsky-Moriya interaction has no influence and we can expect in
this case the well known results of domain wall motion without
Dzyaloshinsky-Moriya interaction.   
\begin{figure}[h]
\includegraphics*[bb = 50 675 590 775,width=8.cm]{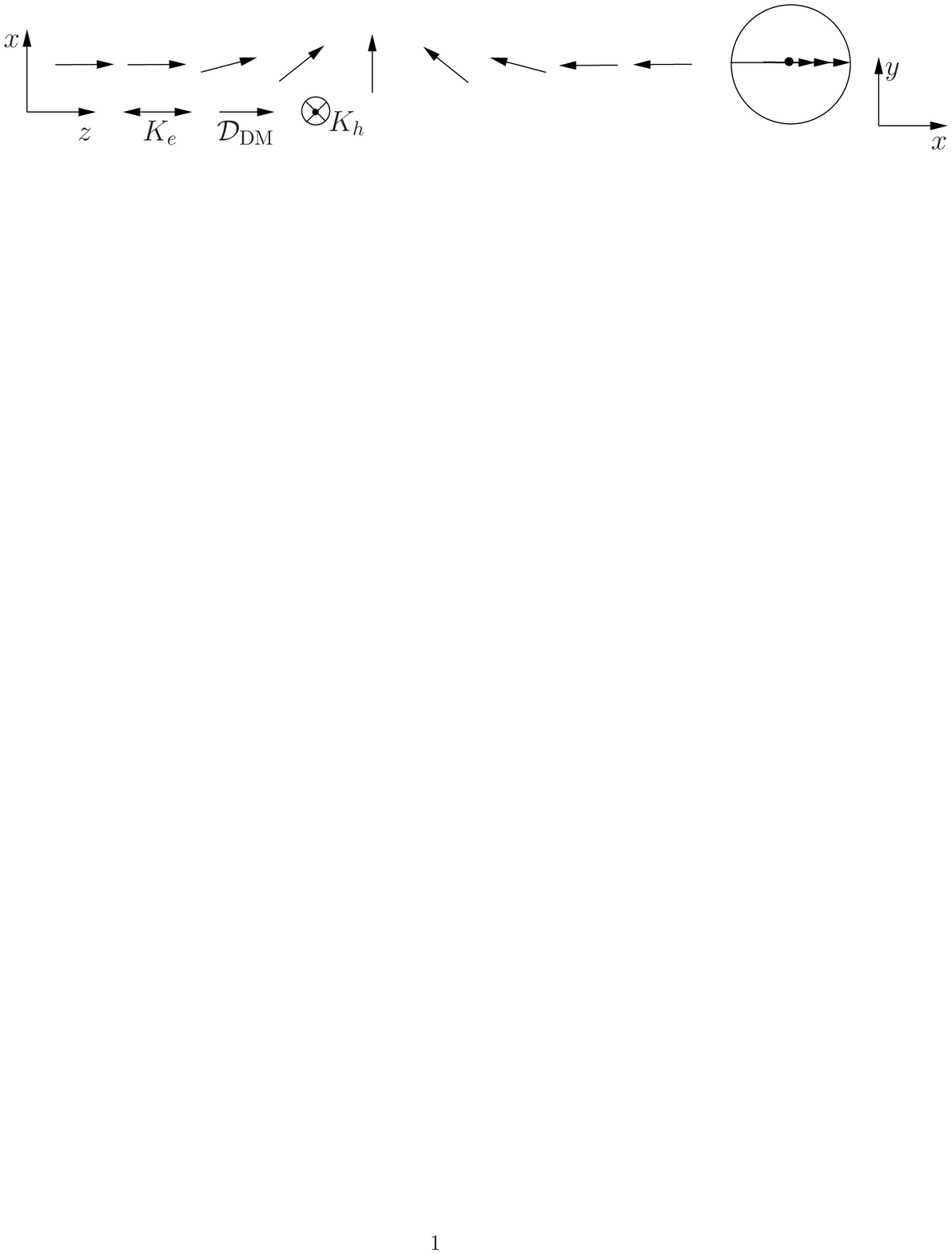}  
  \caption {Sketch of the domain wall profile of a 1D transverse
    domain wall with Dzyaloshinsky-Moriya interaction corresponding to
    the scenario described in Sec.~\ref{S1D}: left hand side
    $x-z$ plane, right hand side (circle) $x-y$ plane. The
    Dzyaloshinsky-Moriya vector  
    ${\bf{\cal D}}_{\mathrm{DM}}$ is oriented parallel to the easy
    axis $K_e$ and perpendicular to the hard axis anisotropy
    $K_h$. Due to the hard axis anisotropy $K_h$ we can assume a
    direct reversal $\mathrm{d}\phi/\mathrm{d}t = 0$.}       
  \label{f:pic1}
\end{figure}

It is quite easy to verify that in this case $\delta {\cal
  E}/\delta \theta = 0$  if we assume $B_z= 0$ (static
solution) for this moment leads to the well known domain wall profile
(\ref{thetaMouginDMmotion}), with the domain wall width: 
\begin{equation} \label{haDW}
\Delta = \sqrt{\frac{A}{K_e+K_h\sin^2\phi}} \;.
\end{equation} 

Then, with $\delta {\cal E}/{\delta \phi}$ under the assumption to be in the
center of the domain wall: $\cos \theta = 0 \Rightarrow \theta = \pi/2$:
\begin{equation}
\left.\frac{\delta {\cal E}}{\delta \phi}\right|_{\theta =
  \frac{\pi}{2}} = K_h \sin(2\phi)\;, 
\end{equation} 
the equations
(\ref{Goofy1}) and (\ref{Goofy2}) become: 
\begin{eqnarray} \label{C1E1}
\frac{v}{\Delta} =
\frac{K_h\gamma}{M_S} \sin(2\phi)  +  
\frac{u_z}{\Delta}  \;,
\end{eqnarray}
and
\begin{eqnarray} \label{C1E2}
0 =  - \gamma B_z+
\frac{\alpha v}{\Delta} - \frac{\beta u_z}{\Delta}\;. 
\end{eqnarray}
Here, the domain wall width $\Delta$ is given by Eq.~(\ref{haDW}).

Inserting Eq.~(\ref{C1E2}) in (\ref{C1E1}) leads to the 
condition which describes the stability of the domain wall (Walker
breakdown): 
\begin{equation} \label{StabilityPhiCondition}
\phi = \frac{1}{2}\mathrm{arcsin}\left(\frac{M_SB_z+ (\beta -
  \alpha)\frac{u_zM_S}{\gamma\Delta}}{\alpha K_h}\right) \;.
\end{equation}
Inserting this result in Eq.~(\ref{haDW}), gives the domain wall
width $\Delta$ during the domain wall motion e.g. $B_z \neq 0$. Furthermore,
Eq.~(\ref{C1E1}) together with 
Eq.~(\ref{StabilityPhiCondition}) leads to the well known velocity 
equation for a domain wall motion with direct reversal
\cite{wieserPRB04,wieserPRB10ii,mouginEPL07}:
\begin{equation} \label{DWVelocityDirekt}
v = \frac{\gamma B_z}{\alpha}\Delta + \frac{\beta}{\alpha}u_z\;.
\end{equation}
The first term describes the influence of an external field and the
second term the influence of the electric current. The
same result can be derived also directly from Eq.~(\ref{C1E2}).

As discussed before the dominating hard axis anisotropy $K_h$ leads to
the fact that the Dzyaloshinsky-Moriya interaction has no
influence. This can be seen also in the equation for the 
velocity Eq.~(\ref{DWVelocityDirekt}) and stability equation
(\ref{StabilityPhiCondition}). The equations for the velocity as
well as the stability of the domain wall are the same as in this
scenario. For comparison, the domain wall 
motion of a transverse domain wall without influence of a
Dzyaloshinsky-Moriya interaction is described in
\cite{wieserPRB10ii,mouginEPL07}.  

On the other hand the assumption of a dominating hard axis anisotropy
$K_h$ is a quite hard criteria which was necessary to solve this
problem analytical. A softer assumption: $K_h$ not dominating will
lead to an influence of the Dzyaloshinsky-Moriya interaction. However,
in this case a complete 
analytical description without approximations is impossible, e.g. Tretiakov
and Abanov \cite{tretiakovPRL10} have used a perturbation theory to
obtain their result.   \\

\subsection{Precessional motion} \label{S1P}
While $K_h \neq 0$ is not exactly solvable without the assumption of
a dominating hard axis anisotropy the situation changes
if we assume $K_h = 0$ (no hard axis anisotropy). As before in
previous subsection we assume the Dzyaloshinsky-Moriya vector parallel
to the easy axis anisotropy in $+z$ direction. Furthermore, we assume
$K_h = 0$, therefore we have a rotational symmetry around the
$z$ axis and this means we cannot neglect the influence of the
Dzyaloshinsky-Moriya interaction: 
$\mathrm{d}\phi/\mathrm{d}z \neq 0$. Furthermore, we have to expect
that the magnetic moments and therefore the domain wall precess during
the motion: $\mathrm{d}\phi/\mathrm{d}t \neq 0$.  
\begin{figure}[h]
\includegraphics*[bb = 50 670 590 775,width=8.cm]{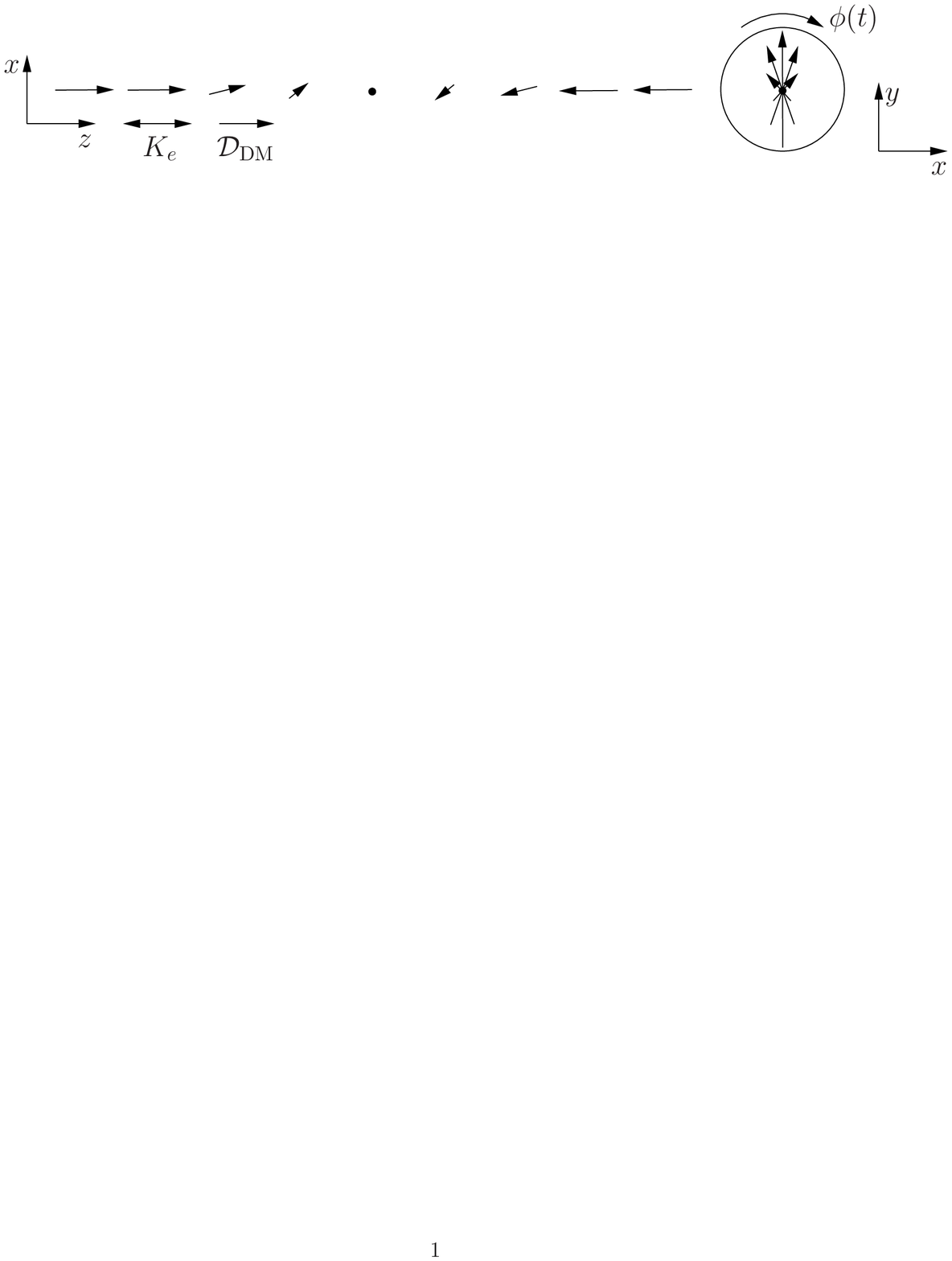}  
  \caption {Sketch of the domain wall profile of a 1D transverse
    domain wall with Dzyaloshinsky-Moriya interaction corresponding to
    the scenario described in Sec.~\ref{S1P}: left hand side
    $x-z$ plane, right hand side (circle) $x-y$ plane. The
    Dzyaloshinsky-Moriya vector 
    ${\bf{\cal D}}_{\mathrm{DM}}$ is oriented parallel to the easy
    axis anisotropy $K_e$. Due to the Dzyaloshinsky-Moriya interaction and the
    absence of the hard axis anisotropy $K_h = 0$ the domain wall profile
    shows a twist and we have to assume a precessional motion
    $\mathrm{d}\phi/\mathrm{d}t \neq 0$.}        
  \label{f:pic2}
\end{figure}

In this scenario $\delta {\cal E}/\delta \phi = 0$ leads to:
\begin{equation} \label{varEphiTwist}
\frac{\delta {\cal E}}{\delta \phi} =
-\frac{\mathrm{d}}{\mathrm{d}z}\left[\sin^2\theta
  \left(2A\frac{\mathrm{d}\phi}{\mathrm{d}z} \pm {\cal
    D}_{\mathrm{DM}}\right)\right] = 0\;,
\end{equation}
which means:
\begin{equation} \label{dphidz}
\frac{\mathrm{d}\phi}{\mathrm{d}z} = \mp \frac{{\cal
    D}_{\mathrm{DM}}}{2A} \;.
 \end{equation}
The solution of this differential equation is given by:
\begin{equation}
\phi(z,t) = \mp \Gamma (z-q(t)) + \phi_0 \;,
\end{equation}
where we have set $z_0 = q$ and $\Gamma = {\cal D}_{\mathrm{DM}}/2A$.
This equation describes the linear increase / decrease of $\phi$ with
$z$ due to the Dzyaloshinsky-Moriya interaction. In other 
words we find a twist in the domain wall profile:
$\mathrm{d}\phi/\mathrm{d}z \neq 0$ (see 
Fig.~\ref{f:pic2}). 

With this knowledge it is easy to show that $\delta {\cal
E}/\delta \theta = 0$, with $B_z= 0$, 
leads to the well known domain wall profile
(\ref{thetaMouginDMmotion}) together with the domain wall width:
\begin{equation}
\Delta = \sqrt{\frac{A}{K_e- \frac{{\cal D}_{\mathrm{DM}}^2}{4A}}} \;.
\end{equation} 

To describe the dynamics it is necessary to know $\delta {\cal
E}/\delta \phi$ and $\mathrm{d}\phi/\mathrm{d}z$: $\delta {\cal
E}/\delta \phi$ is equal to zero [see Eq.~(\ref{varEphiTwist})] and
$\mathrm{d}\phi/\mathrm{d}z$ is described by Eq.~(\ref{dphidz}). 
With this information we can write down our main equations
Eq.~(\ref{Goofy1}) and (\ref{Goofy2}) as: 
\begin{eqnarray} 
\frac{v}{\Delta} &=&  - \alpha\frac{\mathrm{d}\phi}{\mathrm{d}t} +
\frac{u_z}{\Delta} \pm \beta u_z \Gamma  \label{G1DWDMmotion} \\ 
\frac{\mathrm{d}\phi}{\mathrm{d}t} &=& \frac{\alpha v}{\Delta} -
\gamma B_z\pm u_z \Gamma - \frac{\beta u_z}{\Delta} \;. \label{G2DWDMmotion} 
\end{eqnarray}

Inserting (\ref{G2DWDMmotion}) in (\ref{G1DWDMmotion}) we get:
\begin{equation} \label{VFinalDWOscDMz}
v = \frac{\gamma B_z}{\alpha+\frac{1}{\alpha}}\Delta + \frac{1+\alpha\beta}{1+
\alpha^2}u_z \mp \frac{\alpha-\beta}{1+\alpha^2} \Gamma \Delta u_z\;.
\end{equation}
This result is identical to the velocity of a field and current driven
domain wall without Dzyaloshinsky-Moriya interaction
\cite{wieserPRB10ii,mouginEPL07}, just with the
additional term $\mp (\alpha-\beta)/(1+\alpha^2) \Gamma \Delta u_z$
describing the influence of the Dzyaloshinsky-Moriya
interaction. This term increases / decreases the velocity depending on
the direction of the Dzyaloshinsky-Moriya vector. 
  
Inserting (\ref{G1DWDMmotion}) in (\ref{G2DWDMmotion}) leads to:
\begin{equation} \label{PhiDotMoving}
\frac{\mathrm{d}\phi}{\mathrm{d}t} =
\frac{\alpha-\beta}{1+\alpha^2}\frac{u_z}{\Delta} -  
\frac{\gamma B_z}{1+\alpha^2} \pm
\frac{\left(1+\alpha\beta\right)\Gamma}{1+\alpha^2}u_z \;.
\end{equation}
This equation describes the precession of the transverse component of
the domain wall during the motion. 

At this point it should be noticed that we have assumed to be in the
center of the domain wall. This means that we are in a moving
frame. In the case of the domain wall velocity this is not a
problem, because the velocity of the domain wall and the velocity
of the moving frame are identical. However, due to the twisted shape
of the domain wall not only the domain wall rotates during the 
motion but also our coordinate system. This rotation comes from the
twisted shape but does not take into account the precessional motion
of the magnetic moments. In other words Eq.~(\ref{PhiDotMoving}) leads
to wrong results because we are not in a stationary frame or in
the rotating frame with the correct rotational frequency. In the later
case we can expect $\mathrm{d}\phi/\mathrm{d}t = 0$. So, if
we are interested in 
investigating the oscillation of the domain wall we have to be in a
stationary or rotating frame with the same rotational
frequency as the magnetic moments. However, both is not the
case. Therefore, we have to make an additional transformation to get
in one of these two frames. In the 
following we use the stationary frame. 

We know from Eq.~(\ref{dphidz}) that:
\begin{equation}
\frac{\mathrm{d} \phi}{\mathrm{d} z} = \mp \Gamma \hspace{3mm}\Leftrightarrow 
\hspace{3mm} \mathrm{d} \phi = \mp \Gamma \mathrm{d} z\;,
\end{equation}
and therefore:
\begin{equation}
\frac{\mathrm{d} \phi}{\mathrm{d} t} = \mp \Gamma
\frac{\mathrm{d}z}{\mathrm{d} t} \;. 
\end{equation}
This is the correction we have to take into account to change from the
rotating to the stationary frame. $v = \mathrm{d}z/\mathrm{d}t$ is the
velocity of the domain wall and at the same time the velocity of the
moving frame. Adding this term to  
Eq.~(\ref{PhiDotMoving}) and taking into account
Eq.~(\ref{VFinalDWOscDMz}), the precession in the
stationary frame is described by:
\begin{equation}
\frac{\mathrm{d} \phi}{\mathrm{d} t} =
\frac{\alpha-\beta}{1+\alpha^2}\left(\frac{1}{\Delta} \pm \Gamma^2  
\Delta\right)\! u_z - \frac{\gamma B_z}{1+\alpha^2}\left(1 \pm
\alpha\Gamma\Delta  \right)\;.
\end{equation}
Together with equation (\ref{VFinalDWOscDMz}) this equation describes
the precessional domain wall motion of transverse domain walls with a
Dzyaloshinsky-Moriya interaction 
where the Dzyaloshinsky-Moriya vector is parallel to the easy axis
anisotropy. This result is 
identical with the result given by Tretiakov and Abanov
\cite{tretiakovPRL10}, however the given description here is simpler.

\section{Dzyaloshinsky-Moriya vector perpendicular to the easy axis anisotropy
  directions} \label{s:Perp} 
The assumption of the previous section was that the
Dzyaloshinsky-Moriya vector is 
parallel to the easy axis anisotropy. In principle this not the
general situation. Theoretically, the Dzyaloshinsky-Moriya
vector can show in any 
direction, depending on the symmetry of the system. To simplify the
problem and due to the fact that we are interested in analytical
solvable problems we restrict ourself to 
scenarios with the Dzyaloshinsky-Moriya vector perpendicular to the easy axis
anisotropy. In the following, we investigate three scenarios: (1) the
Dzyaloshinsky-Moriya vector 
has an orientation perpendicular to the easy axis anisotropy $K_e$ and
parallel to the hard axis anisotropy $K_h$, (2) the
Dzyaloshinsky-Moriya vector is 
perpendicular to the easy as well hard axis anisotropy $K_h$, and (3)
there is just an easy axis anisotropy ($K_h = 0$). In all these scenarios we
assume that the spin chain has an alignment along the $z$ axis, the
easy axis is oriented in $\pm z$ direction (transverse domain wall)
and the Dzyaloshinsky-Moriya vector is oriented in $y$ direction. To
distinguish the three 
scenarios we change the strength (1),(2) $K_h \neq 0$, (3) $K_h
= 0$ and orientation of the hard axis anisotropy $K_h$: (1) the hard
axis anisotropy $K_h$ is oriented in $\pm y$ direction, (2) in $\pm x$
direction.  

The micromagnetic energy density $\cal E$ in the case of a
Dzyaloshinsky-Moriya vector in $y$ direction parallel to the hard axis
anisotropy $K_h$ is given by:  
\begin{eqnarray} 
{\cal E} &=& A \left[\left(\frac{\mathrm{d} \theta}{\mathrm{d} z}  
\right)^2 + \sin^2\theta \left(\frac{\mathrm{d} \phi}{\mathrm{d}
  z}\right)^2\right]  + {\cal D}_{\mathrm{DM}} \cos \phi
\frac{\mathrm{d} \theta}{\mathrm{d} z}  \nonumber \\ 
&-& M_S B_z\cos \theta + K_h \sin^2\theta \sin^2\phi
- K_e \cos^2\theta   \;,
\end{eqnarray}
For a hard axis anisotropy $K_h$ in $\pm x$ direction we have to
replace the term $K_h \sin^2\theta \sin^2\phi$ by $K_h \sin^2\theta
\cos^2\phi$. \\

\subsection{Direct Reversal: Dzyaloshinsky-Moriya vector parallel to
  $K_h$}  \label{S2D} As before in the first scenario, we assume the
appearance of both uniaxial anisotropies $K_e$ and $K_h$. However, this
time we assume that the Dzyaloshinsky-Moriya vector ${\cal
  D}_{\mathrm{DM}}$ shows in $+y$ direction (parallel to $K_h$) and not in
$+x$ direction. In this case we don't need to assume a dominating hard
axis anisotropy to solve the problem exactly.
 
In this scenario the variation $\delta {\cal E}/\delta \theta$,
with $B_z= 0$, leads to the following result:
\begin{eqnarray}
\frac{\delta {\cal E}}{\delta \theta} &=& 2A 
\frac{\mathrm{d}^2\theta}{\mathrm{d}z^2} - {\cal
  D}_{\mathrm{DM}} \sin\phi \frac{\mathrm{d}\phi}{\mathrm{d}z} \\ 
&+& 2 \sin\theta \cos\theta
\left[K_e + K_h \sin^2\phi - A
  \left(\frac{\mathrm{d}\phi}{\mathrm{d}z}\right)^2 \right] = 0 \nonumber
\end{eqnarray} 
With the assumption that $\phi$ is constant: $\mathrm{d}\phi/\mathrm{d} z
= 0$ and $\mathrm{d}\phi/\mathrm{d} t = 0$ this differential equation can
be solved easily and leads again to the well known domain wall profile
(\ref{thetaMouginDMmotion}) together with an domain wall width which is not
influenced by the Dzyaloshinsky-Moriya interaction:
\begin{equation}
\Delta = \sqrt{\frac{A}{K_e+K_h\sin^2\phi}} \;. 
\end{equation}
\begin{figure}[h]
\includegraphics*[bb = 50 675 590 775,width=8.cm]{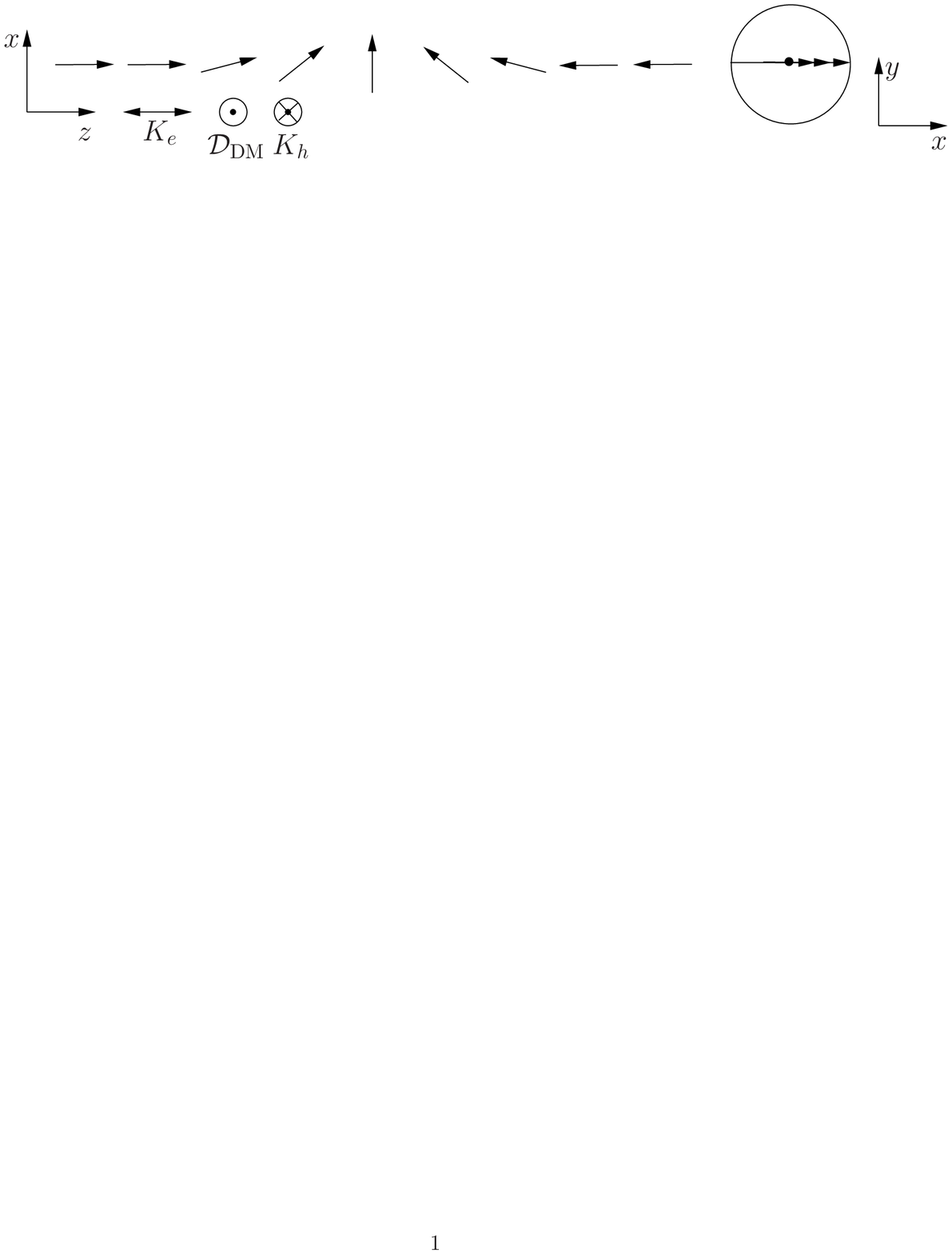}  
  \caption {Sketch of the domain wall profile of a 1D transverse
    domain wall with Dzyaloshinsky-Moriya interaction corresponding to
    the scenario described in Sec.~\ref{S2D}: left hand side
    $x-z$ plane, right hand side (circle) $x-y$ plane. The
    Dzyaloshinsky-Moriya vector 
    ${\bf{\cal D}}_{\mathrm{DM}}$ is oriented perpendicular to the easy
    axis $K_e$ and parallel to the hard axis anisotropy $K_h$. The
    Dzyaloshinsky-Moriya interaction only affects the domain wall
    width but not the profile itself. Due to the hard axis anisotropy
    $K_h$ we can assume a direct reversal $\mathrm{d}\phi/\mathrm{d}t = 0$.}   
  \label{f:pic3}
\end{figure}
The second variation $\delta {\cal E}/\delta \phi$, under the
assumption to be in the center of the domain wall: $\theta = \pi/2$
and a constant $\phi$, leads to:
\begin{equation}
\frac{\delta {\cal E}}{\delta \phi} = \left(2K_h \cos\phi +
\frac{{\cal D}_{\mathrm{DM}}}{\Delta}\right)\sin\phi \;. 
\end{equation}
Therefore, we are able to write the main equations Eq.~(\ref{Goofy1}) and
(\ref{Goofy2}) as:
\begin{eqnarray} \label{uhh}
\frac{v}{\Delta} &=&  \frac{\gamma}{M_S}\left(2K_h \cos\phi +
\frac{{\cal D}_{\mathrm{DM}}}{\Delta} \right)\sin\phi + \frac{u_z}{\Delta}  
\label{Thiaville1} \\ 
0 &=& \frac{\alpha v}{\Delta} - \gamma B_z- \frac{\beta u_z}{\Delta} \;. 
\label{Thiaville2} 
\end{eqnarray}
After eliminating $v$ we get the stability condition:
\begin{equation} \label{WBCondThiaville}
\sin \phi = \frac{\gamma B_z+ (\beta -
  \alpha)\frac{u_z}{\Delta}}{\frac{\alpha \gamma}{M_S}\left(2K_h \cos\phi +
\frac{{\cal D}_{\mathrm{DM}}}{\Delta} \right)} 
\;. 
\end{equation}
Inserting this result in (\ref{uhh}) gives the formula for the
velocity:
\begin{equation}
v = \frac{\gamma B_z}{\alpha}\Delta + \frac{\beta}{\alpha}u_z \;.
\end{equation}
This result is identical to the velocity of a transverse domain wall with
direct reversal and no Dzyaloshinsky-Moriya interaction
\cite{wieserPRB10ii,mouginEPL07}. 
The only difference between both descriptions is the stability 
criteria (\ref{WBCondThiaville}). In the case with
Dzyaloshinsky-Moriya interaction the domain walls are more stable,
which means the Walker breakdown appears at higher field and current values. \\

\subsection{Direct Reversal: Dzyaloshinsky-Moriya vector perpendicular to
  $K_h$} \label{S3D}
\begin{figure}[h]
\includegraphics*[bb = 50 670 590 775,width=8.cm]{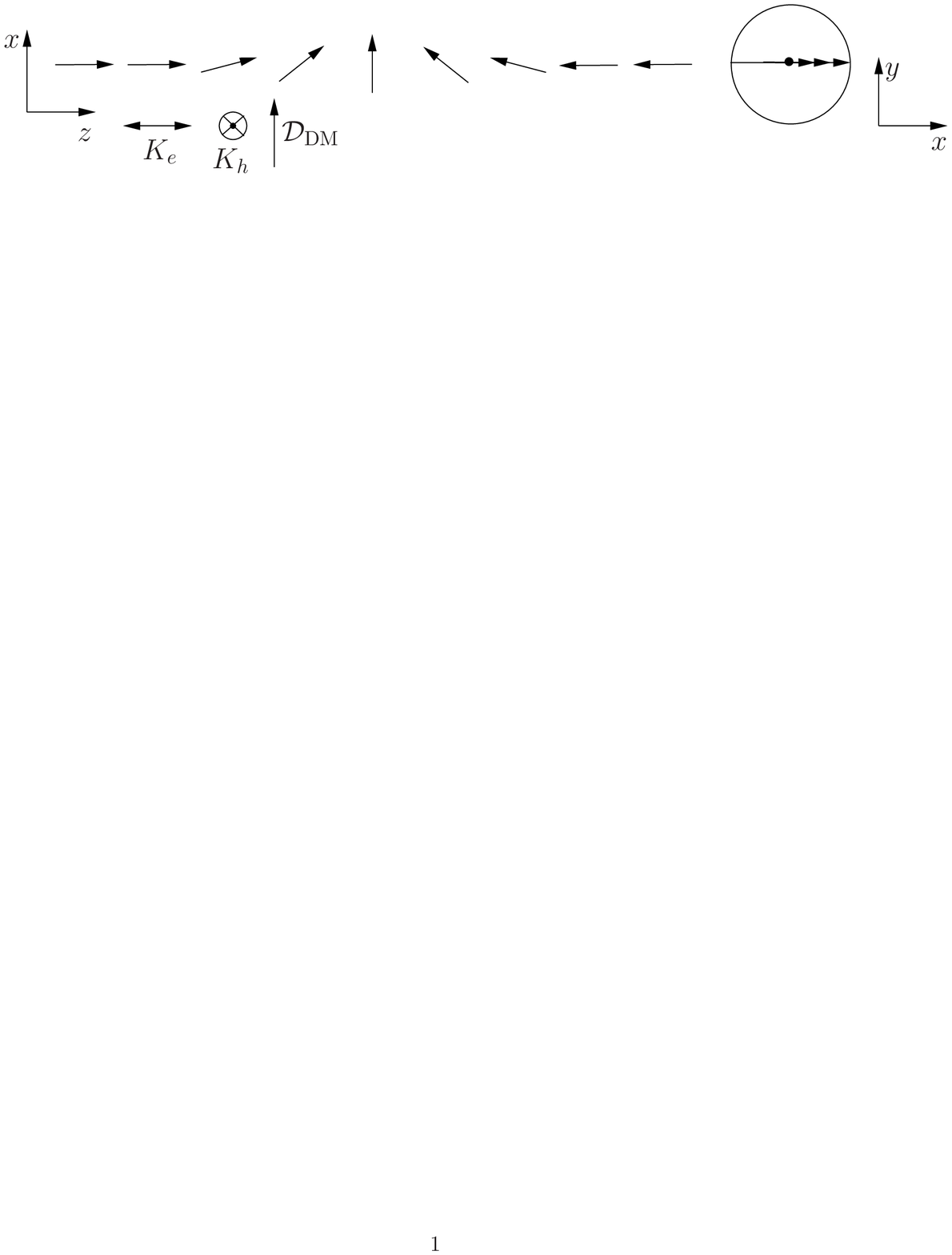}    
  \caption {Sketch of the domain wall profile of a 1D transverse
    domain wall with Dzyaloshinsky-Moriya interaction corresponding to
    the scenario described in Sec.~\ref{S3D}: left hand side
    $x-z$ plane, right hand side (circle) $x-y$ plane. The
    Dzyaloshinsky-Moriya vector 
    ${\bf{\cal D}}_{\mathrm{DM}}$ is oriented perpendicular to the easy
    axis $K_e$ and hard axis anisotropy $K_h$. The
    Dzyaloshinsky-Moriya interaction shows no influence on the
    profile in this case. Due to the hard axis anisotropy $K_h$ we
    can assume a direct reversal $\mathrm{d}\phi/\mathrm{d}t = 0$.}       
  \label{f:pic4}
\end{figure}

In the following we assume the same geometry as before, however with
the following difference: the Dzyaloshinsky-Moriya vector is now
oriented in $+x$ direction and therefore perpendicular to both
anisotropies $K_e$ and $K_h$ (see Fig.~\ref{f:pic4}).

This scenario is solvable if we assume that both anisotropies are
dominating with respect to Dzyaloshinsky-Moriya interaction: $K_h \gg
{\cal D}_{\mathrm{DM}}$, 
$K_e \gg {\cal D}_{\mathrm{DM}}$. 
In this case the Dzyaloshinsky-Moriya interaction can be
neglected and we find a normal transverse domain wall with the profile
(\ref{thetaMouginDMmotion}) and the domain wall width:
\begin{equation}
\Delta = \sqrt{\frac{A}{K_e+K_h\sin^2\phi}} \;. 
\end{equation} 
Then, the stability is characterized by:
\begin{equation} \label{StabilityPhiCondition2}
\phi = \frac{1}{2}\mathrm{arcsin}\left(\frac{M_SB_z+ (\beta -
  \alpha)\frac{u_zM_S}{\gamma\Delta}}{\alpha K_h}\right) \;.
\end{equation}
and the velocity of the domain wall described by:
\begin{equation} \label{DWVelocityDirekt}
v = \frac{\gamma B_z}{\alpha}\Delta + \frac{\beta}{\alpha}u_z\;.
\end{equation}
\\

\subsection{Precessional motion} \label{S2P} 
\begin{figure}[h]
\includegraphics*[bb = 50 665 590 775,width=8.cm]{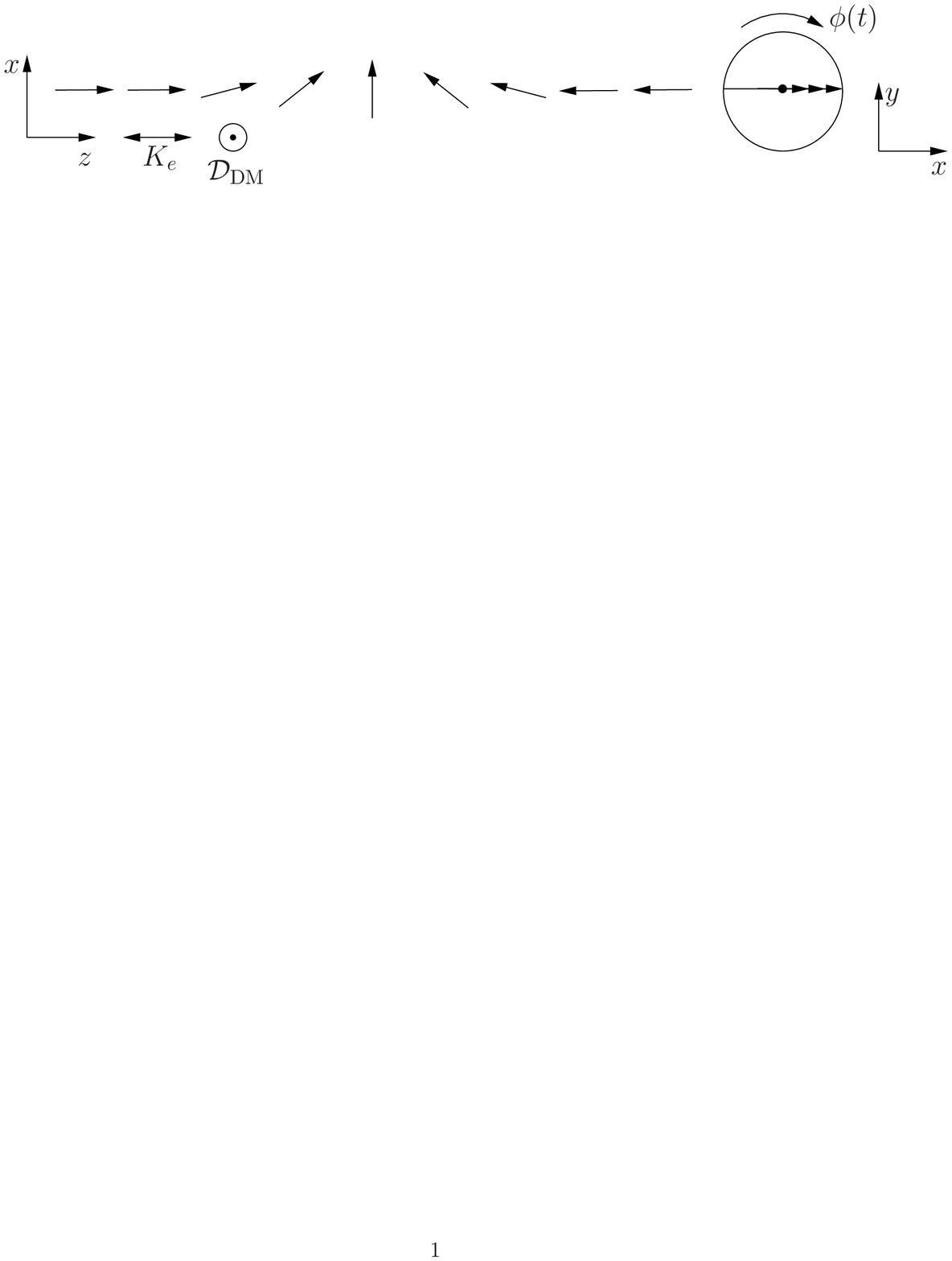}    
  \caption {Sketch of the domain wall profile of a 1D transverse
    domain wall with Dzyaloshinsky-Moriya interaction corresponding to
    the scenario described in Sec.~\ref{S2P}: left hand side
    $x-z$ plane, right hand side (circle) $x-y$ plane. The
    Dzyaloshinsky-Moriya vector 
    ${\bf{\cal D}}_{\mathrm{DM}}$ is oriented perpendicular to the easy
    axis anisotropy $K_e$. Due to the absence of the hard axis
    anisotropy $K_h$ we can assume a precessional motion
    $\mathrm{d}\phi/\mathrm{d}t \neq 0$ which leads to a time
    dependent domain wall width $\Delta(t)$ and domain wall energy
    $E(t)$.
}       
  \label{f:pic5}
\end{figure}

In this subsection we assume $K_h = 0$, which means no 
hard axis anisotropy and therefore a precessional motion $\mathrm{d}
\phi/\mathrm{d}t \neq 0$. The Dzyaloshinsky-Moriya vector in this
scenario is oriented in $+y$ direction (see Fig.~\ref{f:pic5}). 
Furthermore, for simplicity we assume that $\mathrm{d}
\phi/\mathrm{d}z = 0$. This is definitively the case when the
transversal component of the domain wall and the Dzyaloshinsky-Moriya vector
$\vec{D}_{\mathrm{DM}}$ are perpendicular as in
Fig.~\ref{f:pic5}. In this case the Dzyaloshinsky-Moriya interaction
assists the magnetization  
reversal in the $x-z$ plane which is described by the angle $\theta$. The
assumption $\mathrm{d} \phi/\mathrm{d}z = 0$ is not necessarily the
case when $\vec{D}_{\mathrm{DM}}$ is parallel to the transversal
component of the domain wall. 

Under these assumptions, especially $\mathrm{d} \phi/\mathrm{d}z =
0$, we find:
\begin{eqnarray}
\frac{\delta {\cal E}}{\delta \theta} =
2A\frac{\mathrm{d}^2\theta}{\mathrm{d}z^2} + 2 K_e \sin\theta\cos\theta  
= 0 \;,
\end{eqnarray} 
and therefore, the well know domain wall profile
(\ref{thetaMouginDMmotion}) together with the domain wall width:
\begin{equation}
\Delta = \sqrt{\frac{A}{K_e}} \;, 
\end{equation}
and the domain wall energy:
\begin{equation}
E = 4\sqrt{A K_e} + \pi \cal{D}_{\mathrm{DM}}\;.  
\end{equation}

The variation $\delta {\cal E}/\delta \phi$ under the assumption to be
in the center of the 
domain wall $(\theta = \pi/2)$ leads to:
\begin{equation}
\left.\frac{\delta {\cal E}}{\delta \phi}\right|_{\theta = \pi/2} = 
\frac{{\cal D}_{\mathrm{DM}}}{\Delta}\sin\phi \;. 
\end{equation}
Then, it is easy to write the equations (\ref{Goofy1}) and
(\ref{Goofy2}) as:
\begin{equation} \label{arg}
\frac{v}{\Delta} =  \alpha \frac{\mathrm{d}\phi}{\mathrm{d}t} -
\frac{\gamma}{M_S}\frac{{\cal D}_{\mathrm{DM}}}{\Delta}\sin\phi 
+ \frac{u_z}{\Delta}  \;, 
\end{equation}
and 
\begin{equation} \label{urg}
\frac{\mathrm{d}\phi}{\mathrm{d}t} = \frac{\alpha v}{\Delta} -\gamma
B_z - \frac{\beta u_z}{\Delta} \;.
\end{equation} 
Eliminating $\mathrm{d}\phi/\mathrm{d}t$ leads to the following
velocity equation:
\begin{equation}
v(t) = \frac{\gamma B_z}{\alpha+\frac{1}{\alpha}}\Delta +
\frac{1+\alpha\beta}{1+\alpha^2}u_z  - \frac{\gamma {\cal D}_{\mathrm{DM}}
  \sin\phi(t)}{M_S \Delta (1+\alpha^2)}\;.
\end{equation} 

In time average (mean value) the velocity of the domain wall is equal to: 
\begin{equation} \label{vaverage}
\overline{v} = \frac{\gamma B_z}{\alpha+\frac{1}{\alpha}}\Delta +
\frac{1+\alpha\beta}{1+\alpha^2}u_z \;,
\end{equation}
which is equal to the velocity of a transverse domain wall with
precession and without Dzyaloshinsky-Moriya
interaction \cite{wieserPRB10ii}. 
\begin{figure}[h]
\includegraphics*[bb = 124 530 460 770,width=7.cm]{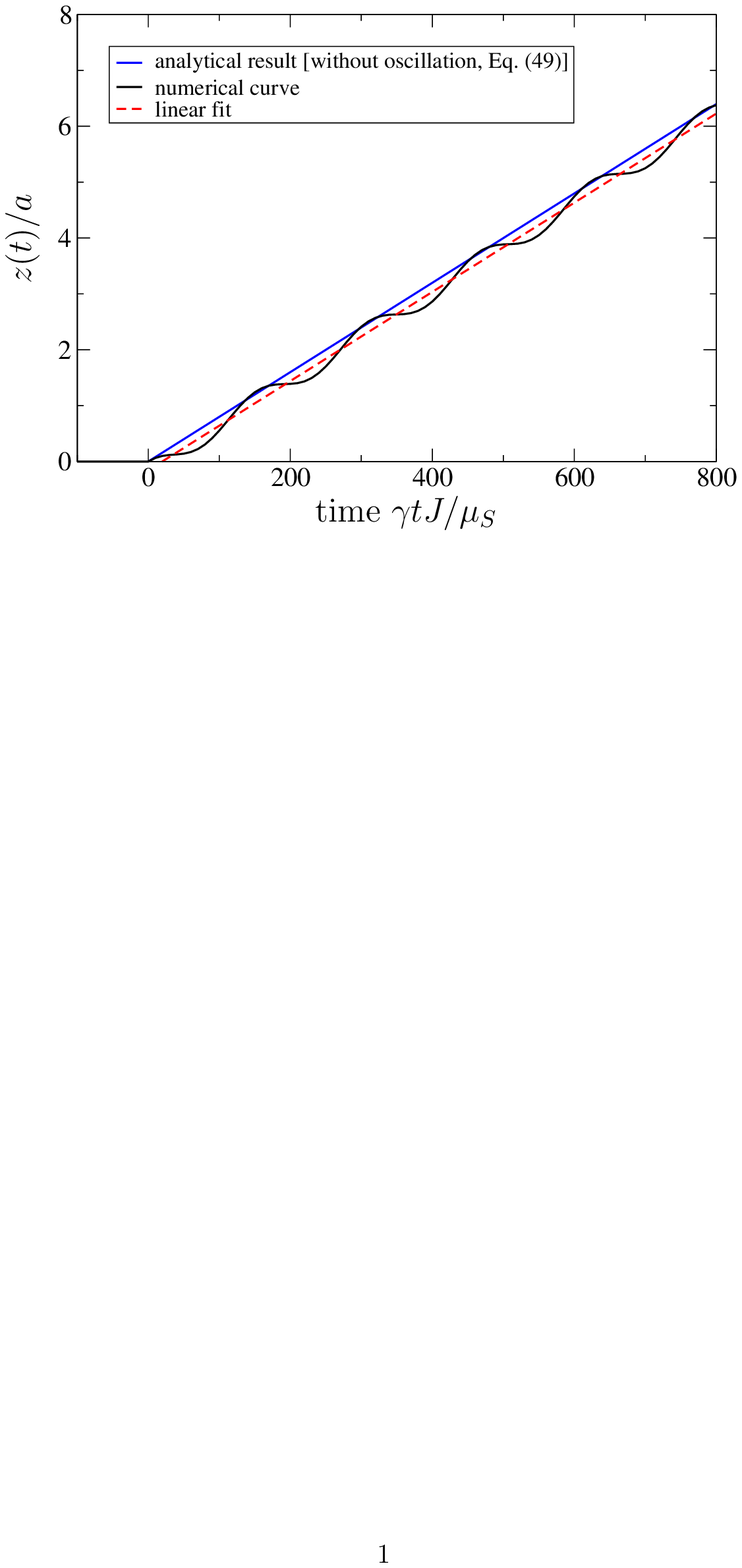}    
  \caption {Domain wall motion of a transverse domain wall with
    Dzyaloshinsky-Moriya interaction corresponding to
    the scenario described in Sec.~\ref{S2P}: domain wall position as function
  of time corresponding to the scenario that the Dzyaloshinsky-Moriya
  vector ${\bf{\cal 
      D}}_{\mathrm{DM}}$ shows an orientation perpendicular to the easy
    axis anisotropy $K_e$ and the assumption that there is no
    additional hard axis anisotropy $K_h = 0$. The last assumption leads
  to a precessional motion which leads to the oscillation in
  $z(t)$. The curves coming from computer simulations with: $\mu_SB_z/J =
  0.2$, $D_{\mathrm{DM}}/J = 0.004$, and $D_e/J = 0.005$.}       
  \label{f:pic6}
\end{figure}

Eliminating $v$ in Eq.~(\ref{arg}) and (\ref{urg}) lead to the following
differential equation for $\phi$:
\begin{equation} \label{DGLThiavilleOsc}
\frac{\mathrm{d}\phi}{\mathrm{d}t} =
\frac{\alpha-\beta}{1+\alpha^2}\frac{u_z}{\Delta} - 
\frac{\gamma B_z}{1+\alpha^2} -  \frac{\alpha\gamma {\cal
    D}_{\mathrm{DM}}}{M_S \Delta (1+\alpha^2)} \sin\phi(t)\;.
\end{equation}   
This differential equation is needed to be solved to calculate the
velocity $v(t)$. For simplification we write the differential equation
Eq.~(\ref{DGLThiavilleOsc}) as:   
\begin{equation} \label{DGLThiavilleOsc2}
\frac{\mathrm{d}\phi}{\mathrm{d}t} = \phi_0 - \phi_1 \sin\phi(t)  \;,
\end{equation}   
with 
\begin{equation}
\phi_0 = \frac{\alpha-\beta}{1+\alpha^2}\frac{u_z}{\Delta} -
\frac{\gamma B_z}{1+\alpha^2} \;, 
\end{equation}
and
\begin{equation}
\phi_1 =  \frac{\alpha\gamma {\cal D}_{\mathrm{DM}}}{M_S
  \Delta (1+\alpha^2)}\;. 
\end{equation}
Eq.~(\ref{DGLThiavilleOsc2}) can be easily solved by separation of
variables. The result is: 
\begin{equation}
\phi(t) = 2
\mathrm{arctan}\left[\frac{\sqrt{\phi_0^2-\phi_1^2}}{\phi_1}\tan\left(\frac{ 
\sqrt{\phi_0^2-\phi_1^2}t}{2}\right)-\frac{\phi_0}{\phi_1}\right] \;.
\end{equation}

An interesting situation appears if the domain wall is driven by
an electrical current $(u_z \neq 0, B_z = 0)$ and if we assume $\alpha
= \beta$. It is known that in this case a transverse domain wall without
Dzyaloshinsky-Moriya interaction shows no
precession $\mathrm{d}\phi/\mathrm{d}t = 0$ even if the domain wall normally
precess $\mathrm{d}\phi/\mathrm{d}t \neq 0$ if $\alpha \neq
\beta$. However, in the case with Dzyaloshinsky-Moriya interaction
we find a precessional motion even in this special situation: $\alpha
= \beta$, $B_z = 0$. In this case we find:
\begin{equation}
\phi(t) = 2
\mathrm{arctan}\left[\mathrm{exp}\left(-\frac{\alpha\gamma {\cal
    D}_{\mathrm{DM}}}{M_S \Delta (1+\alpha^2)}t\right)\right] \;,
\end{equation}
and
\begin{equation}
\overline{v} = u_z \;.
\end{equation}

To get a better feeling about the accuracy and correctness of the
results of the last subsection additional numerical simulations have
been performed. Therefore the Gilbert equation (\ref{GEqS}) has been solved
using the discrete Heisenberg model (\ref{Heisenberg}) and as
numerical solver the Heun method. Starting point
of the simulation was a relaxed domain wall in the first half of a
linear chain of the length of 1000 magnetic moments. The orientation
of the chain is along the $z$ axis and the magnetic momets inside the
domains are in $\pm z$ direction. The unrelaxed domain wall is a
$180^\circ$ head-to-head transverse domain wall which becomes the final
shape during the relaxation process. After switching on
an electric current the domain wall displacement has been determined
by looking for the zero-crossing of the $\tanh$-profile of the domain
wall: $S_z = 0$. For more details of the computer simulations on the
atomic length scale please see
\cite{wieserPRB04,schiebackEPJB07,wieserPRB10ii}. 
 
Fig.~\ref{f:pic6} shows the domain wall displacement as function of
time for a current driven transverse domain wall described as in the
last scenario with $\alpha < \beta$. As predicted by analytical
calculations the domain wall shows an oscillatory behavior and moves
with the averaged velocity given by Eq.~(\ref{vaverage}). However due 
to the oscillation the domain wall arrives the end of the system with
a delay or earlier depending on the sign of the Dzyaloshinsky-Moriya
vector ${\bf{\cal D}}_{\mathrm{DM}}$.

\section{Summary} \label{s:summary}
In summary this manuscript gives a complete overview of the domain
wall motion of transverse domain walls under the influence of the
Dzyaloshinsky-Moriya interaction. The description is analytical using
the $q-\phi$ model and covers the scenarios of a Dzyaloshinsky-Moriya
vector parallel 
respectively perpendicular to the easy axis direction and the fact
that the domain wall can show a direct reversal or precessional
motion. The first two scenarios describe the situation of a
Dzyaloshinsky-Moriya vector 
parallel to the easy axis direction. These scenarios have been
described by Tretiakov and Abanov \cite{tretiakovPRL10} however using
too complex description. The given description within this paper is
simpler and therefore easier to understand. The third scenario is
identical to the description of Thiaville et
al. \cite{thiavilleEPL12} of a Dzyaloshinsky-Moriya vector
perpendicular to easy axis 
and parallel to the hard axis direction. However, here
we give some more informations. The fourth scenario describes
the situation of a Dzyaloshinsky-Moriya vector perpendicular to the
easy axis and at the 
same time hard axis anisotropy. In this case the Dzyaloshinsky-Moriya
interaction has no influence 
as long we assume that ${\cal D}_{\mathrm{DM}}$ is small with respect
to the anisotropies $K_e$ and $K_h$. The fifth scenario assumes a
Dzyaloshinsky-Moriya 
vector perpendicular to easy axis direction. However, within this
scenario we assume a precessional motion. This scenario has not been
described so far and compliments the description of the dynamics of
transverse domain walls under the influence of a Dzyaloshinsky-Moriya
interaction. We have seen that the Dzyaloshinsky-Moriya interaction
can lead to a modification of the velocity, or at least the stability
condition. Other scenarios show no influence of the
Dzyaloshinsky-Moriya interaction on the velocity. To compliment these
results additional computer simulations could be helpful, especially
in cases where the domain wall cannot be described by a simple
transverse domain wall.

\begin{acknowledgments}
This work has been supported by the Deutsche Forschungsgemeinschaft in the
framework of subproject B3 of the SFB 668 and by the Cluster of Excellence
``Nanospintronics''. 
\end{acknowledgments}

\bibliography{Cite}

\end{document}